\documentclass[aps,twocolumn,prd,showpacs,tightenlines]{revtex4}
\usepackage{amssymb}
\usepackage{epsfig}
\usepackage{amsbsy}
\usepackage{amsmath}
\usepackage{graphicx}

\begin{document}

\title{Recovering Unitarity of Lee Model in Bi-Orthogonal Basis}
\author{T. Shi and C. P. Sun}
\affiliation{Institute of Theoretical Physics, Chinese Academy of Sciences, Beijing,
100190, China }

\begin{abstract}
We study how to recover the unitarity of Lee model with the help of
bi-orthogonal basis approach, when the physical coupling constant in
renormalization exceeds its critical value, so that the Lee's Hamiltonian is
non-Hermitian with respect to the conventional inner product. In a very
natural fashion, our systematic approach based on bi-orthogonal basis leads
to an elegant definition of inner product with a non-trivial metric, which
can overcome all the previous problems in Lee model, such as non-Hermiticity
of the Hamiltonian, the negative norm, the negative probability and the
non-unitarity of the scattering matrix.
\end{abstract}

\maketitle

\section{Introduction}

In 1954, T.D. Lee introduced an exactly soluble model to display the
necessity of renormalization in quantum field theory (QFT) \cite{Lee}. It is
delicate that with the Lee model, the renormalizations of the mass,
wave-function and coupling constant were performed in a closed form \cite%
{Schweber}. It is worth remanding that, in the single excitation subspace,
the Lee model can be reduced into the Fano model in atom-molecular physics
\cite{F} or the Anderson model of on-site interaction free in condensed
matter physics \cite{A}. With this recognition, the quantum manipulation for
the coherent transport in quantum dot array has been investigated based on
the Lee model \cite{Zhou1}.

Originally, the Lee model describes the reaction processes $%
V\rightleftarrows N+\theta $, namely, a fermion $V$ transforms to another
fermion $N$ by emitting bosonic $\theta $-particle, and vice versa. In the
single particle subspace---the $V/N\theta $ sector, the Lee model is exactly
solvable. According to the analytic solution, we only need to consider the
mass renormalization of the $V$-particle and coupling constant
renormalization since the bare single-particle states of $\theta $ and $N$
are both the eigenstates of the Hamiltonian, thus the mass renormalizations
of $\theta $ and $N$-particles are not need, and their physical masses equal
to the bare masses. In the $V/N\theta $ sector, the eigenstates of the
Hamiltonian contain the physical $V$-state and $N\theta $-scattering states.

In the Lee model, the renormalization about $V$-particle results in some
perplexing and interesting natures, e.g., the emergence of the ghost state
with an negative norm. Actually when the physical coupling constant $g$,
which obtained from the standard procedure of renormalization, is strong
enough to exceed a critical coupling constant $g_{c}$, the bare coupling
constant in the original Lee model is imaginary, and then the Hamiltonian
becomes non-Hermitian with respect to the conventional inner product in
quantum mechanics (QM) and QFT. To avoid this non-Hermiticity, one could
restrict $g$ to be smaller than $g_{c}$, but the vanishing critical value $%
g_{c}$ due to infinite cutoff makes $g$ always greater than $g_{c}$. In this
sense, the non-Hermiticity is so intrinsic, that it is ineluctable.

To overcome the ghost state problem due to the non-Hermiticity, a
significant attempt \cite{Redmond,EG-NL} is to construct a
Hermitian, but non-local Hamiltonian corresponding to the modified
$V$-particle-Green's function of the Lee model by eliminating the
ghost pole artificially. It was K\"{a}llen and Pauli \cite{Pauli}
who first introduced an indefinite metric such that the norm of
the physical $V$-state is positive, but the norm of the ghost
state is still negative. In 1968 and 1969, Lee himself and Wick
\cite{LW1,LW2} used the same indefinite metric to discuss the
unitarity of the $S$-matrix in the $N\theta $ and $N\theta \theta
$ sectors for the imaginary physical masses of the physical
$V$-state and ghost state in the regime $g>g_{c}$. Recently,
Bender \textit{et al.} \cite{Bender} introduced a different inner
product by a coupling dependent $\mathcal{CPT}$ transformation to
insure the positive norms of the all eigenstates of the
non-Hermitian Hamiltonian. An equivalent Hamiltonian \cite{Jones}
for the Lee model was found by the similar transformation.

In this paper, we use the bi-orthogonal basis approach \cite{Sun,Bi-O} to
find a non-trivial metric and construct a new inner product for the Lee
model in the strong coupling regime $g>g_{c}$. In a natural and consistent
way, this approach overcome the overall problems in applications of the Lee
model due to the non-Hermiticity of the Hamiltonian, such as the negative
norm and ghost state, the negative probability, and the non-unitarity of the
scattering matrix.

Our approach for the Lee model based on the bi-orthogonal basis is to use
the two complete eigenstate sets $\{|E_{n}\rangle \}$ and $\{|D_{n}\rangle
\} $ of the Lee's Hamiltonian $H$ and its conjugate $H^{\dagger }$. The
non-trivial metric is defined by an operator $\eta $ \cite{Bi-O}: $%
|D_{n}\rangle =\eta |E_{n}\rangle $, which result a new inner product $(\phi
,\varphi )=\langle \phi |\eta |\varphi \rangle $. We can explicitly
calculate the metric operator $\eta $ for both the QM Lee model (or called
one boson mode Lee model where the $\theta $-particle only possesses a
single mode) and the QFT model. Using this new inner product, we show that
in the regime $g>g_{c}$, the Hamiltonian is Hermitian, all eigenstates have
positive norms and the scattering matrix is unitary. It is more important
that, our obtained metric for the Lee model is different from that in Ref.
\cite{Bender}, and automatically insures the orthogonality of the different
eigenstates. In this inner product space the Hermiticity of the Hamiltonian
implies the unitarity of the evolution. Therefore, with arbitrary
interaction strength the QFT Lee model becomes acceptable in physics, and
thus could be applied to practical systems, such as a coupled resonator
array interacting with a two level system \cite{Zhou1}.

The paper is organized as follows: In Sec. II, we reconsider the unitarity
breaking by briefly reviewing the Lee model. In Sec. III, we introduce the
concept of bi-orthogonal basis and use it to deal with a simple model, the
QM\ Lee model, whose Hamiltonian is non-Hermitian with respect to the
conventional inner product. We show find the new metric so that the
Hamiltonian becomes Hermitian with respect to the inner product defined by
this metric. In Sec. IV, we use the bi-orthogonal basis to find the new
metric for the QFT Lee model when $g>g_{c}$ so that the Hamiltonian is
Hermitian and the $S$-matrix is unitary in the Hilbert space with this inner
product. In Sec. V, the results are summarized with some comments. In
Appendix, we discuss the relation between the Lee model and an standard
model in quantum optics and cavity QED.

\section{Breaking of Unitarity of Lee Model}

In this section, we revisit the breaking of unitarity by briefly reviewing
the basic properties of the Lee model in the conventional representation.
Then we show why the bi-orthogonal basis is indeed needed to recovery the
unitarity of the Lee model.

\subsection{Lee model}

To describe the reaction processes $V\rightleftarrows N+\theta $, the Lee
model uses Hamiltonian $H=H_{0}+H_{1}$, where
\begin{equation}
H_{0}=m_{0}V^{\dagger }V+m_{N}N^{\dagger }N+\sum_{k}\omega
_{k}a_{k}^{\dagger }a_{k},
\end{equation}%
and%
\begin{equation}
H_{1}=\sum_{k}\frac{g_{0}f_{k}}{\sqrt{2\omega _{k}\Omega }}(a_{k}V^{\dagger
}N+h.c.).
\end{equation}%
Here, $V$ ($N$) is the annihilation operator of the $V$ ($N$)-particle with
bare mass $m_{0}$ ($m_{N}$). The operator $a_{k}$ ($a_{k}^{\dagger }$) is
the annihilation (creation) operator of the massive $\theta $-particle with
the dispersion relation $\omega _{k}=\sqrt{k^{2}+\mu ^{2}}$ and the mass $%
\mu $. Here, we introduce the effective mode volume $\Omega $, the real
cutoff function $f_{k}$ and the bare coupling constant $g_{0}$. In the
infinite cutoff, namely, the function $f_{k}$ tends to unit. Here, we point
that the Lee model neglects the processes $V+\theta \rightleftarrows N$ when
the condition $m_{0}>m_{N}$ is satisfied.

We focus on the $V/N\theta $ sector, i.e., the subspace with $%
n_{V}+n_{\theta }=1$ spanned by the basis $V^{\dagger }\left\vert
0\right\rangle $ and $N^{\dagger }a_{k}^{\dagger }\left\vert 0\right\rangle $%
, where $n_{V}=V^{\dagger }V$ and $n_{\theta }=\sum_{k}a_{k}^{\dagger }a_{k}$%
. In this subspace, we assume the eigenstates%
\begin{equation}
\left\vert E\right\rangle =[cV^{\dagger }+\sum_{p}\phi _{E}(p)N^{\dagger
}a_{p}^{\dagger }]\left\vert 0\right\rangle ,
\end{equation}%
satisfy the eigen-equation $H\left\vert E\right\rangle =E\left\vert
E\right\rangle $, where $E$ denote the corresponding eigenvalues. It follows
from the eigen-equation that%
\begin{eqnarray}
m_{0}c+\sum_{p}\frac{g_{0}f_{p}}{\sqrt{2\omega _{p}\Omega }}\phi _{E}(p)
&=&Ec,  \label{L1} \\
(E-m_{N}-\omega _{p})\phi _{E}(p) &=&\frac{g_{0}f_{p}}{\sqrt{2\omega
_{p}\Omega }}c,  \label{L2}
\end{eqnarray}%
determines the parameters $c$, the wave-functions $\phi _{E}(p)$ and the
eigen-energies $E$. Non-vanishing solutions of Eqs. (\ref{L1}) and (\ref{L2}%
) lead to the secular equation $h(E)=0$, where%
\begin{equation}
h(E)=E-m_{0}-\sum_{p}\frac{g_{0}^{2}f_{p}^{2}}{2\omega _{p}\Omega }\frac{1}{%
E-m_{N}-\omega _{p}}.  \label{E}
\end{equation}%
If $g_{0}$ is real and $h(m_{N}+\mu )>0$ (see Fig. 1a), there are two kinds
of real solutions of Eq. (\ref{E}), where one solution satisfies $%
E<m_{N}+\mu $, the others are $E_{k}=m_{N}+\omega _{k}$. Here, the condition
$h(m_{N}+\mu )>0$, i.e.,%
\begin{equation}
m_{N}+\mu +\sum_{p}\frac{g_{0}^{2}f_{p}^{2}}{2\omega _{p}\Omega }\frac{1}{%
\omega _{p}-\mu }>m_{0},
\end{equation}%
ensures the existence of the stable $V$-particle \cite{unstable} whose
eigen-energy has not imaginary part (see the appendix).

\begin{figure}[tbp]
\includegraphics[bb=29 516 574 734, width=9 cm, clip]{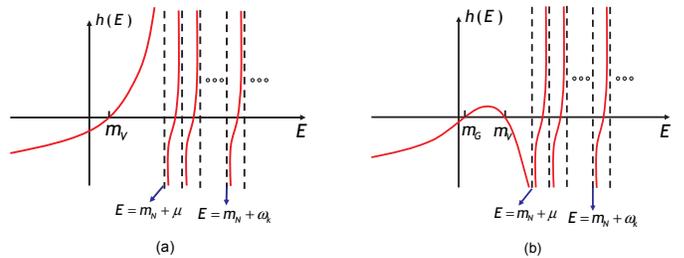}
\caption{(Color online) The schematic for the function $h(E)$: the red
curves denote the function $h(E)$, and the solutions $E=m_{N}+\protect\omega %
_{k}$ are the energy of the scattering states, (a) when $g<g_{c}$, the
equation $h(E)=0$ possesses only one real root $E=m_{V}$ in the regime $%
E<m_{N}+\protect\omega _{k}$ when $h(m_{N}+\protect\mu )>0$; (b) when $%
g>g_{c}$, the equation $h(E)=0$ possesses two real roots $E=m_{V}$ and $%
E=m_{G}$ in the regime $E<m_{N}+\protect\omega _{k}$ when $m_{0}<m_{N}+%
\protect\mu $.}
\end{figure}

We find physical $V$-state with $E=m_{V}$ in the regime $m_{V}<m_{N}+\mu $
as
\begin{equation}
\left\vert \mathbf{V}\right\rangle =Z_{V}^{1/2}[V^{\dagger }+\sum_{p}\phi
_{V}(p)N^{\dagger }a_{p}^{\dagger }]\left\vert 0\right\rangle ,
\end{equation}%
where the momentum representation of wave-function is%
\begin{equation}
\phi _{V}(p)=\frac{g_{0}f_{p}}{\sqrt{2\omega _{p}\Omega }}\frac{1}{%
m_{V}-m_{N}-\omega _{p}},
\end{equation}%
and the normalization constant is%
\begin{equation}
Z_{V}^{-1}=1+\sum_{p}\phi _{V}^{2}(p).  \label{DZ}
\end{equation}%
When $g_{0}\rightarrow 0$, the eigenstate $\left\vert \mathbf{V}%
\right\rangle $ becomes $V^{\dagger }\left\vert 0\right\rangle $. Thus, the
dressed state $\left\vert \mathbf{V}\right\rangle $ describes the
renormalized $V$-particle, or called the physical $V$-particle. The physical
mass of the $V$-particle is $m_{V}$, which satisfies $h(m_{V})=0$.

Next, we consider the scattering states%
\begin{equation}
\left\vert \mathbf{N\theta }_{k}\right\rangle _{\pm }=[c_{k,\pm }V^{\dagger
}+\sum_{p}\phi _{k,\pm }(p)N^{\dagger }a_{p}^{\dagger }]\left\vert
0\right\rangle ,  \label{S}
\end{equation}%
to satisfy $h(E=E_{k})=0$ for $E_{k}=m_{N}+\omega _{k}$. While we assume%
\begin{equation}
c_{k,\pm }=g_{0}f_{k}h^{-1}(m_{N}+\omega _{k}\pm i0^{+})/\sqrt{2\omega
_{k}\Omega },
\end{equation}%
the scattering features are reflected by the $\delta $\emph{-function} in
the wave-function%
\begin{equation}
\phi _{k,\pm }(p)=\delta (k-p)+\frac{g_{0}f_{p}c_{k,\pm }}{\sqrt{2\omega
_{p}\Omega }}\frac{1}{\omega _{k}-\omega _{p}\pm i0^{+}}.
\end{equation}%
The positive infinitesimal $0^{+}$ is introduced here such that the
eigenstates (\ref{S}) are rewritten as the standard Lippmann- Schwinger
scattering states%
\begin{eqnarray}
\left\vert \mathbf{N\theta }_{k}\right\rangle _{\pm } &=&(1+\frac{1}{%
E_{k}-H\pm i0^{+}}H_{I})N^{\dagger }a_{k}^{\dagger }\left\vert
0\right\rangle ,  \label{LS} \\
&=&N^{\dagger }a_{k}^{\dagger }\left\vert 0\right\rangle +\frac{1}{%
E_{k}-H_{0}\pm i0^{+}}H_{I}\left\vert \mathbf{N\theta }_{k}\right\rangle
_{\pm }.  \notag
\end{eqnarray}

Obviously, the eigenstates (\ref{S}) as well as Eq. (\ref{LS}) describe the
scattering of the $\theta $-particle with momentum $k$ by the $N$-particle.
The corresponding $S$-matrix element%
\begin{equation}
S_{pk}=\,_{-}\left\langle \mathbf{N\theta }_{p}\left\vert \mathbf{N\theta }%
_{k}\right\rangle _{+}\right. =e^{2i\delta _{k}}\delta _{kp},
\end{equation}%
is determined by the scattering phase shift $\delta _{k}=\arctan \beta _{k}$%
, where%
\begin{equation}
\beta _{k}=\frac{g_{0}^{2}Z_{V}k}{4\pi (m_{V}-m_{N}-\omega _{k})}[1-\frac{%
g_{0}^{2}Z_{V}}{(2\pi )^{3}}\Theta _{k}]^{-1},  \label{b}
\end{equation}%
and%
\begin{equation}
\Theta _{k}=\mathcal{P}\int \frac{d^{3}p(m_{V}-m_{N}-\omega _{k})}{2\omega
_{p}(\omega _{p}-\omega _{k})(m_{V}-m_{N}-\omega _{p})^{2}}.
\end{equation}%
Here, $\mathcal{P}$ denotes the principal-value integral. Because $%
Z_{V}^{-1} $ in Eq. (\ref{DZ}) contains the divergent integral $\sum_{p}\phi
_{V}^{2}(p) $ when the cutoff tends to infinite, i.e., $f_{p}\rightarrow 1$,
the normalization constant $Z_{V}$ tends to zero so that the phase shift $%
\delta _{k}$ and the cross section both vanish. For avoiding the vanishing
of the cross section, we introduce the physical coupling constant $%
g^{2}=g_{0}^{2}Z_{V}$ and assigne it to be finite.

\subsection{Breaking of Unitarity and Emerging Ghost State}

The normalization constant can be rewritten as%
\begin{equation}
Z_{V}=1-\frac{g^{2}}{g_{c}^{2}},
\end{equation}%
in terms of the critical coupling constant $g_{c}^{2}$ defined as%
\begin{equation}
g_{c}^{-2}=\sum_{p}\frac{f_{p}^{2}}{2\omega _{p}\Omega }\frac{1}{%
(m_{V}-m_{N}-\omega _{p})^{2}}.
\end{equation}%
Then the relation between $g^{2}$ and $g_{0}^{2}$ is%
\begin{equation}
g_{0}^{2}=\frac{g^{2}g_{c}^{2}}{g_{c}^{2}-g^{2}}\text{.}
\end{equation}%
Obviously, if $g^{2}<g_{c}^{2}$, the normalization constant $Z_{V}$ and the
square $g_{0}^{2}$ of the bare coupling constant are always positive, so
that the conventional approach based on QFT is self-consistent and proper.
However, if $g^{2}>g_{c}^{2}$, the normalization constant is negative,
namely, the physical $V$-state has the negative norm. In addition, the
square $g_{0}^{2}$ of bare coupling constant becomes negative, so that $%
g_{0}=i\lambda _{0}$ ($\lambda _{0}\in $ real number) is imaginary, which
results in that the Lee model Hamiltonian is not Hermitian with respect to
the conventional inner product and the unitarity is broken.

As illustrated in Fig. 1b, when $g_{0}$ is imaginary, the detail analysis
shows that when $m_{0}>m_{N}+\mu $, there is no real solution satisfied $%
E<m_{N}+\mu $, but when $m_{0}<m_{N}+\mu $ there are two real solutions
satisfied $E<m_{N}+\mu $. In this paper, we consider the case $%
m_{0}<m_{N}+\mu $ and prove that if $g_{0}$ is imaginary, the secular
equation $h(E)=0$ possesses another real solution $E=m_{G}$ ($<m_{V}$) in
the regime $E<m_{N}+\mu $ besides the real root $E=m_{V}$.

To consider the physics of this solution we rewrite $h(E)$ as $%
h(E)=Z_{V}^{-1}(E-m_{V})F(E)$, where%
\begin{equation}
F(E)=1-\frac{g^{2}}{(2\pi )^{3}}\int \frac{d^{3}p(E-m_{V})}{2\omega
_{p}\Lambda (E,\omega _{p})\Lambda ^{2}(m_{V},\omega _{p})}.
\end{equation}%
and $\Lambda (E,\omega _{p})=E-m_{N}-\omega _{p}$. The schematics of the
functions $h(E)$ and $F(E)$ are shown in Fig. 1a (b) and Fig. 2a (b) for the
case $g>g_{c}$ ($g<g_{c}$), which display the relations $\partial _{E}F(E)>0$
and $F(m_{V})=1$ explicitly.

\begin{figure}[tbp]
\includegraphics[bb=48 318 563 577, width=8 cm, clip]{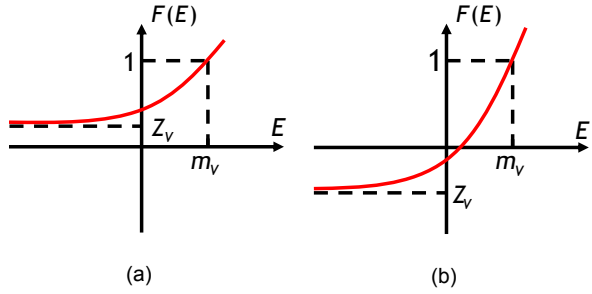}
\caption{(Color online) The schematic for the function $F(E)$: the red
curves denote the function $F(E)$, (a) when $g<g_{c}$, the equation $F(E)=0$
possesses no root in the regime $E<m_{N}+\protect\omega _{k}$; (b) when $%
g>g_{c}$, the equation $F(E)=0$ possesses one real root $E=m_{G}$ in the
regime $E<m_{N}+\protect\omega _{k}$.}
\end{figure}
It is remarkable that as $E\rightarrow -\infty $, $F(E)\rightarrow Z_{V}$.
Then we conclude that if $g>g_{c}$, i.e., $Z_{V}<0$, there always exists the
real root $m_{G}$ of equation $h(E)$ in the interval $(-\infty ,m_{V})$ when
$m_{0}<m_{N}+\mu $, which is the energy of the ghost state. Here, the
un-normalized ghost state is%
\begin{equation}
\left\vert \mathbf{G}\right\rangle =[V^{\dagger }+\sum_{p}\phi
_{G}(p)N^{\dagger }a_{p}^{\dagger }]\left\vert 0\right\rangle ,
\end{equation}%
where wave-function in the momentum representation is%
\begin{equation}
\phi _{G}(p)=\frac{g_{0}f_{p}}{\sqrt{2\omega _{p}\Omega }}\frac{1}{%
m_{G}-m_{N}-\omega _{p}}.
\end{equation}%
The energy spectrums of the Lee model in the $V/N\theta $ sector are
displayed in Fig. 3 schematically for the case $g>g_{c}$ and $g<g_{c}$.

To overcome the difficulty of negative norm when $g>g_{c}$, the metric $%
\zeta =(-1)^{n_{V}}$ is introduced by K\"{a}llen and Pauli \cite%
{Pauli,LW1,LW2}.\ They found that for the case $g>g_{c}$, in the new inner
product space the norms $\left\langle \mathbf{V}\right\vert \zeta \left\vert
\mathbf{V}\right\rangle $ and $_{\pm }\left\langle \mathbf{N\theta }%
_{k}\right\vert \zeta \left\vert \mathbf{N\theta }_{k}\right\rangle _{\pm }$
were both positive, and the orthogonal relations $_{\pm }\left\langle
\mathbf{N\theta }_{k}\right\vert \zeta \left\vert \mathbf{V}\right\rangle
=\left\langle \mathbf{V}\right\vert \zeta \left\vert \mathbf{N\theta }%
_{k}\right\rangle _{\pm }=0$ were also ensured. However, the norm $%
\left\langle \mathbf{G}\right\vert \zeta \left\vert \mathbf{G}\right\rangle $
of the ghost state was still negative and thus the ghost state phenomenon
still exists! The negative norm of the ghost state implies that the metric $%
\zeta $\ is an indefinite metric, and the $S$-matrix satisfies $\zeta
S^{\dagger }\zeta S=1$, which is non-unitary. Recently, Bender \textit{et al}%
. \cite{Bender} introduced the different inner product by the $\mathcal{CPT}$
symmetry to insure the positive norms of the all eigenstates in the sector $%
V/N\theta $ for $g>g_{c}$. In this paper, we use the standard
bi-orthogonal basis approach to find the proper inner product that
automatically ensures the orthogonality and the positive definite
of the eigenstates.
\begin{figure}[tbp]
\includegraphics[bb=49 236 558 632, width=8.5 cm, clip]{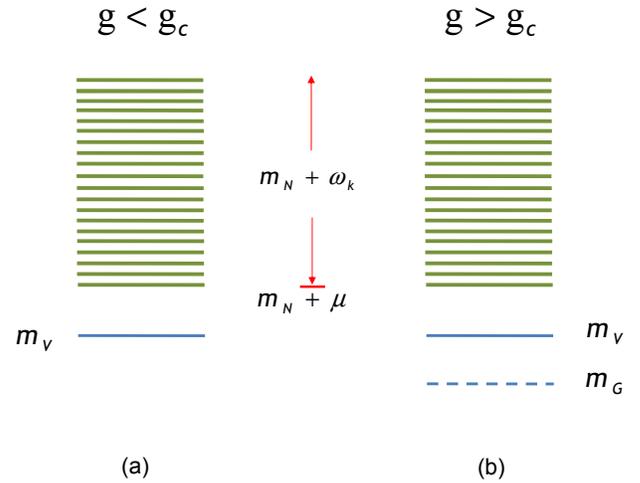}
\caption{(Color online) The schematic for the energy spectrum in the $V/N%
\protect\theta $ sector: when $g>g_{c}$, the ghost state emerges as a
discrete state labelled by $m_{G}$. The discrete state labelled by $m_{V}$
is the physical $V$-state. Here, $m_{N}+\protect\omega _{k}$ denotes the
energy of the scattering state. $m_{V}$ and $m_{G}$ denote the energies of
the physical $V$-particle state and ghost state. }
\end{figure}

\section{Bi-orthogonal basis used for one boson mode Lee model}

In this section, we first briefly introduce the key ideas of the
bi-orthogonal basis with its application to the one boson Lee model (or
called QM Lee model) where the $\theta $-particle only has one mode, which
has been studied extensively \cite{Bender,Jones,Trubatch}.

\subsection{Bi-orthogonal basis approach and Metric operator}

In the bi-orthogonal basis approach to the non-Hermitian Hamiltonian $H$,
the eigenstates and the corresponding eigenvalues $\left\vert
e_{n}\right\rangle $ and $E_{n}$ determined by the eigen-equation $%
H\left\vert e_{n}\right\rangle =E_{n}\left\vert e_{n}\right\rangle $ can
span the whole Hilbert space in some sense, but arbitrary two eigenstates
are usually not orthogonal to each other. To have a complete orthogonal
basis, we need the eigenstates $\left\vert d_{m}\right\rangle $ of $%
H^{\dagger }$ with corresponding eigenvalues $E_{m}^{D}$. Then we have the
two sets of the basis \{$\left\vert e_{n}\right\rangle $\} and \{$\left\vert
d_{m}\right\rangle $\}, and we can prove that if $E_{m}^{D}\neq E_{n}^{\ast
} $ $\left\langle d_{m}\left\vert e_{n}\right\rangle \right. =0$, and if $%
E_{m}^{D}=E_{n}^{\ast }$ $\left\langle d_{m}\left\vert e_{n}\right\rangle
\right. $ is nonzero. Hereafter, we define $\left\vert d_{n}\right\rangle $
is the eigenstate with the eigenvalue $E_{n}^{\ast }$ of the Hamiltonian $%
H^{\dagger }$ and do not assume their normalization. The two sets of the
basis \{$\left\vert e_{n}\right\rangle $\} and \{$\left\vert
d_{n}\right\rangle $\} form the so-called bi-orthogonal basis \cite{Sun,Bi-O}%
.

With the help of bi-orthogonal basis, the completeness relation is%
\begin{equation}
\sum_{n}\frac{\left\vert e_{n}\right\rangle \left\langle d_{n}\right\vert }{%
\left\langle d_{n}\left\vert e_{n}\right\rangle \right. }=\sum_{n}\frac{%
\left\vert d_{n}\right\rangle \left\langle e_{n}\right\vert }{\left\langle
e_{n}\left\vert d_{n}\right\rangle \right. }=1.
\end{equation}%
If we define the normalized basis by $\left\vert E_{n}\right\rangle
=\left\vert e_{n}\right\rangle /\sqrt{\left\langle d_{n}\left\vert
e_{n}\right\rangle \right. }$ and $\left\vert d_{n}\right\rangle =\left\vert
d_{n}\right\rangle /\sqrt{\left\langle e_{n}\left\vert d_{n}\right\rangle
\right. }$, we have the generic completeness relations%
\begin{equation}
\sum |E_{n}\rangle \langle D_{n}|=\sum |D_{n}\rangle \langle E_{n}|=1,
\end{equation}%
and
\begin{equation}
\langle D_{n}|E_{m}\rangle =\langle E_{n}|D_{m}\rangle =\delta _{mn}.
\end{equation}

According to Refs. \cite{Bi-O}, we can define the new inner product
\begin{equation}
(\Phi ,\Psi )=\left\langle \Phi ^{D}\left\vert \Psi \right\rangle
=\left\langle \Phi \right\vert \eta \left\vert \Psi \right\rangle \right. ,
\end{equation}%
with new metric operator $\eta $:%
\begin{equation}
|E_{n}\rangle \rightarrow |D_{n}\rangle =\eta |E_{n}\rangle .
\end{equation}%
The Hamiltonian $H$ under the new metric is Hermitian, i.e., $\eta
H=H^{\dagger }\eta $ or $(\Phi ,H\Psi )^{\ast }=(\Psi ,H\Phi )$. Thus, for
the non-Hermitian Hamiltonian that satisfies $\eta H=H^{\dagger }\eta $, we
always find out the metric $\eta $ by bi-orthogonal basis, so that in the
new inner product space the Hamiltonian becomes Hermitian and the all
eigenstates have the positive norms, and the arbitrary two eigenstates are
orthogonal to each other.

\subsection{One boson mode Lee model}

In this subsection, we use the bi-orthogonal basis to deal with the simple
model, i.e., one boson mode Lee model or called QM Lee model.\ To show the
main ideas to solve the unitarity problem of the standard Lee model (the QFT
Lee Model), we first consider the one boson mode Lee model, namely, the $%
\theta $-particle only has one mode, which has been studied extensively \cite%
{Bender,Jones,Trubatch}. It can be proved that this model is equivalent to
the \textquotedblleft standard model" of quantum optics, the Jaynes-Cummings
(JC) model \cite{JC} (see the appendix).

The one boson mode Lee model is described by the Hamiltonian $%
H_{s}=H_{0}+H_{1}$, where%
\begin{equation}
H_{0}=m_{0}V^{\dagger }V+m_{N}N^{\dagger }N+\varepsilon a^{\dagger }a,
\end{equation}%
and%
\begin{equation}
H_{1}=g_{0}(aV^{\dagger }N+h.c.).
\end{equation}%
When $g_{0}=i\lambda _{0}$ is imaginary, we use the bi-orthogonal basis
approach to reconsider the Hermiticity of $H_{s}$. The eigenstates of the
Hamiltonian $H_{s}$ in the subspace spanned by the basis $V^{\dagger
}\left\vert 0\right\rangle $ and $N^{\dagger }a^{\dagger }\left\vert
0\right\rangle $ are%
\begin{equation}
\left\vert \Psi _{\pm }\right\rangle =\mathcal{N}_{\pm }^{-1}[i\lambda
_{0}V^{\dagger }+(E_{\pm }-m_{0})N^{\dagger }a^{\dagger }]\left\vert
0\right\rangle ,
\end{equation}%
and the corresponding eigenvalues are%
\begin{equation}
E_{\pm }=\frac{1}{2}(s\pm \sqrt{u^{2}-4\lambda _{0}^{2}}),
\end{equation}%
where the normalization constants $\mathcal{N}_{\pm }$ are determined below.
Here, $s=m_{N}+\varepsilon +m_{0}$ and $u=m_{N}+\varepsilon -m_{0}$.
Obviously, the eigenvalues are all real only if $u^{2}-4\lambda _{0}^{2}>0$.
Here, we focus on the case $u^{2}-4\lambda _{0}^{2}>0$ so that all energies
are real in this sector. Below we only consider the case $u>2\lambda _{0}$,
and the other case $u<-2\lambda _{0}$ can be discussed in the similar manner.

It is clear that with respect to the conventional inner product $%
\left\langle \Phi \left\vert \Psi \right\rangle \right. $ the two
eigenstates are not orthogonal to each other, i.e., $\left\langle \Psi _{\mp
}\left\vert \Psi _{\pm }\right\rangle \right. \neq 0$, and the normalization
constants are%
\begin{equation}
\mathcal{N}_{\pm ,\mathrm{old}}^{2}=u(E_{\pm }-m_{0}),  \label{N}
\end{equation}%
where the subscript \textquotedblleft $\mathrm{old}$\textquotedblright\
denotes the normalization constants are defined in the conventional inner
product space. In the dual representation the eigenstates of the conjugate
Hamiltonian $H_{s}^{\dagger }=H_{0}-H_{1}$ are%
\begin{equation}
\left\vert \Psi _{\pm }^{D}\right\rangle =\pm \mathcal{N}_{\pm
}^{-1}[-i\lambda _{0}V^{\dagger }+(E_{\pm }-m_{0})N^{\dagger }a^{\dagger
}]\left\vert 0\right\rangle ,
\end{equation}%
with the corresponding eigenvalues $E_{\pm }^{D}=E_{\pm }$. It can be
verified that $\left. \left\langle \Psi _{+}^{D}\right\vert \Psi
_{-}\right\rangle =\left. \left\langle \Psi _{-}^{D}\right\vert \Psi
_{+}\right\rangle =0$, and $\left. \left\langle \Psi _{\pm }^{D}\right\vert
\Psi _{\pm }\right\rangle \neq 0$.

To overcome the non-orthogonality difficulty, we find out the new metric%
\begin{equation}
\eta =\left\vert \Psi _{+}^{D}\right\rangle \left\langle \Psi
_{+}^{D}\right\vert +\left\vert \Psi _{-}^{D}\right\rangle \left\langle \Psi
_{-}^{D}\right\vert ,
\end{equation}%
or its explicit form in operators:%
\begin{eqnarray}
\eta &=&\frac{1}{\sqrt{u^{2}-4\lambda _{0}^{2}}}[u(V^{\dagger }V+N^{\dagger
}N)  \notag \\
&&-2i\lambda _{0}(V^{\dagger }Na-h.c.)],  \label{Yi}
\end{eqnarray}%
by solving the equation $\left\vert \Psi _{\pm }^{D}\right\rangle =\eta
\left\vert \Psi _{\pm }\right\rangle $. Then we define the new inner product
$(\Phi ,\Psi )=\left\langle \Phi \right\vert \eta \left\vert \Psi
\right\rangle =\left\langle \Phi ^{D}\right\vert \Psi \rangle =\left\langle
\Phi \right\vert \Psi ^{D}\rangle $: $|\Psi ^{D}\rangle =\eta \left\vert
\Psi \right\rangle $ in the single particle subspace. The orthogonality $%
\left. \left\langle \Psi _{+}^{D}\right\vert \Psi _{-}\right\rangle =\left.
\left\langle \Psi _{-}^{D}\right\vert \Psi _{+}\right\rangle =0$ and the
positively definite norm of the eigenstates are both ensured in the new
inner product space defined by metric $\eta $. And the Hamiltonian is also
Hermitian, i.e., $\eta H=H^{\dagger }\eta $. The new normalization condition
$\left. \left\langle \Psi _{+}^{D}\right\vert \Psi _{+}\right\rangle =\left.
\left\langle \Psi _{-}^{D}\right\vert \Psi _{-}\right\rangle =1$ gives%
\begin{equation}
\mathcal{N}_{\pm }^{2}=\sqrt{u^{2}-4\lambda _{0}^{2}}(E_{\pm }-m_{0})>0,
\end{equation}%
which are different from the normalization constants (\ref{N}) determined by
the conventional inner product. Then we also have the completeness relation%
\begin{eqnarray}
\eta (\left\vert \Psi _{+}\right\rangle \left\langle \Psi _{+}\right\vert
+\left\vert \Psi _{-}\right\rangle \left\langle \Psi _{-}\right\vert ) &=&1,
\\
(\left\vert \Psi _{+}\right\rangle \left\langle \Psi _{+}\right\vert
+\left\vert \Psi _{-}\right\rangle \left\langle \Psi _{-}\right\vert )\eta
&=&1.
\end{eqnarray}

\section{Lee model with $g>g_{c}$}

In this section, we study the QFT Lee model by using the bi-orthogonal basis
approach. The problem is considered in the subspace with $n_{\theta
}+n_{V}=1 $ when the bare coupling constant $g_{0}=i\lambda _{0}$ ($\lambda
_{0}\in R$) is imaginary, i.e., $g>g_{c}$. We will prove in the appendix
that the QFT Lee model is essentially a massive JC model where the
\textquotedblleft light field\textquotedblright\ would be massive.

We first consider the ghost state and physical $V$-state. We show that in
the new inner product space, the ghost state and physical $V$-state have
positive norms. Secondly, we consider the scattering states of the $\theta $%
-particle. We prove that the scattering states also have positive norms in
the new inner product space.

Let me first present the new metric operator%
\begin{eqnarray}
\eta &=&bV^{\dagger }V+\sum_{kp}W_{kp}N^{\dagger }Na_{k}^{\dagger }a_{p}
\notag \\
&&+\sum_{k}(s_{k}V^{\dagger }Na_{k}+h.c.),  \label{yi}
\end{eqnarray}%
\ for the Lee model as our central result.\ Here, the parameters $b=2\Xi
_{G}-1$ and $s_{p}=2\Xi _{G}\phi _{G}(p)$ are defined by%
\begin{eqnarray}
\Xi _{G}^{-1} &=&1+\sum_{p}\phi _{G}^{2}(p),  \label{nor} \\
W_{kp} &=&\delta _{kp}-2\Xi _{G}\phi _{G}(k)\phi _{G}(p).
\end{eqnarray}

It will be proved soon as follows that the inner product $(\Phi ,\Psi
)=\left\langle \Phi \right\vert \eta \left\vert \Psi \right\rangle $ based
on this metric will solve all the problems in the Lee model with $g>g_{c}$:
(1) the ghost state and physical $V$-state have positive norms; (2) the
scattering states of the $\theta $-particle has the positive norms; (3) the
orthogonality of the eigenstates is ensured; (4) the unitarity of the $S$%
-matrix for the $N\theta \rightarrow N\theta $ process is recovered
automatically.

\subsection{Ghost state and physical $V$-state}

The energies of the ghost state and physical $V$-state are located out of
the continuum, i.e., $m_{L}-m_{N}-\omega _{k}<0$ for $L=\mathbf{V}$ and $%
\mathbf{G}$. In this case, the ghost state and physical $V$-state are
expressed as an unified form%
\begin{equation}
\left\vert L\right\rangle =\Xi _{L}^{1/2}[V^{\dagger }+\sum_{p}\phi
_{L}(p)N^{\dagger }a_{p}^{\dagger }]\left\vert 0\right\rangle .  \label{L}
\end{equation}%
Here, the normalized constants $\Xi _{L}$ are determined by the
bi-orthogonal basis approach below. To find out the proper inner product, we
consider that two dual eigenstates
\begin{eqnarray}
\left\vert \mathbf{V}^{D}\right\rangle &=&\Xi _{V}^{1/2}[-V^{\dagger
}+\sum_{p}\phi _{V}(p)N^{\dagger }a_{p}^{\dagger }]\left\vert 0\right\rangle
,  \notag \\
\left\vert \mathbf{G}^{D}\right\rangle &=&\Xi _{G}^{1/2}[V^{\dagger
}-\sum_{p}\phi _{G}(p)N^{\dagger }a_{p}^{\dagger }]\left\vert 0\right\rangle
,  \label{LD}
\end{eqnarray}%
of $H^{\dagger }$ have the same eigenvalues $m_{L}$. Then we define the
inner products $(\mathbf{V},\mathbf{V})=\left\langle \mathbf{V}%
^{D}\left\vert \mathbf{V}\right\rangle \right. $, $(\mathbf{G},\mathbf{G}%
)=\left\langle \mathbf{G}^{D}\left\vert \mathbf{G}\right\rangle \right. $, $(%
\mathbf{V},\mathbf{G})=\left\langle \mathbf{V}^{D}\left\vert \mathbf{G}%
\right\rangle \right. $ and $(\mathbf{G},\mathbf{V})=\left\langle \mathbf{G}%
^{D}\left\vert \mathbf{V}\right\rangle \right. $ for the ghost state and
physical $V$-state. The normalization conditions $\left\langle \mathbf{V}%
^{D}\left\vert \mathbf{V}\right\rangle \right. =\left\langle \mathbf{G}%
^{D}\left\vert \mathbf{G}\right\rangle \right. =1$ lead to the constants $%
\Xi _{G}^{-1}$ in Eq. (\ref{nor}) and%
\begin{equation}
\Xi _{V}=\frac{g^{2}}{g_{c}}-1>0.
\end{equation}%
Obviously, with respect to the new inner products, the conditions $(\mathbf{V%
},\mathbf{V})=(\mathbf{G},\mathbf{G})=1$ and $(\mathbf{V},\mathbf{G})=(%
\mathbf{G},\mathbf{V})=0$ are satisfied simultaneously.

\subsection{Scattering states and unitarity of the $S$-matrix}

The energies of the scattering states are $E_{k}=m_{N}+\omega _{k}$. The
corresponding eigenstates of Hamiltonian $H$ are $\left\vert \mathbf{N\theta
}_{k}\right\rangle _{\pm }$. In the dual space, the corresponding eigenstates%
\begin{equation}
\left\vert \mathbf{N\theta }_{k}^{D}\right\rangle _{\pm }=[-c_{k,\pm
}V^{\dagger }+\sum_{p}\phi _{k,\pm }(p)N^{\dagger }a_{p}^{\dagger
}]\left\vert 0\right\rangle ,  \label{NtD}
\end{equation}%
of Hamiltonian $H^{\dagger }$ have the same eigenvalues $E_{k}=m_{N}+\omega
_{k}$. The eigenstates (\ref{NtD}) are rewritten as%
\begin{equation}
\left\vert \mathbf{N\theta }_{k}^{D}\right\rangle _{\pm }=N^{\dagger
}a_{k}^{\dagger }\left\vert 0\right\rangle +\frac{1}{E_{k}-H_{0}\pm i0^{+}}%
H_{I}^{\dagger }\left\vert \mathbf{N\theta }_{k}^{D}\right\rangle _{\pm }\,,
\label{LS2}
\end{equation}%
which are the standard form of\ the Lippmann-Schwinger represntation.
Obviously, the eigenstates (\ref{LS}), (\ref{L}), (\ref{LD}), and (\ref{LS2}%
)\ immediately give the orthogonal relations%
\begin{eqnarray}
_{\pm }\left\langle \mathbf{N\theta }_{k}^{D}\left\vert \mathbf{N\theta }%
_{p}\right\rangle _{\pm }\right. &=&\,_{\pm }\left\langle \mathbf{N\theta }%
_{p}\left\vert \mathbf{N\theta }_{k}^{D}\right\rangle _{\pm }\right. =\delta
_{kp},  \notag \\
\left\langle \mathbf{V}^{D}\left\vert \mathbf{N\theta }_{p}\right\rangle
_{\pm }\right. &=&\,_{\pm }\left\langle \mathbf{N\theta }_{p}^{D}\left\vert
\mathbf{V}\right\rangle \right. =0.
\end{eqnarray}

With the help of the eigenstates of the Hamiltonian $H$ and the dual
Hamiltonian $H^{\dagger }$, we find the new metric $\eta $, which is
satisfies $\eta \left\vert \Psi \right\rangle =\left\vert \Psi
^{D}\right\rangle $, so that the inner product is $(\Phi ,\Psi
)=\left\langle \Phi \right\vert \eta \left\vert \Psi \right\rangle $. For
the eigenstates, the relations $\eta \left\vert \mathbf{V}\right\rangle
=\left\vert \mathbf{V}^{D}\right\rangle $, $\eta \left\vert \mathbf{G}%
\right\rangle =\left\vert \mathbf{G}^{D}\right\rangle $, and $\eta
\left\vert \mathbf{N\theta }_{k}\right\rangle _{\pm }=\left\vert \mathbf{%
N\theta }_{k}^{D}\right\rangle _{\pm }$ immediately result in the metric (%
\ref{yi}). Under this proper metric, in this subspace all states always have
the positive norms, and the arbitrary two eigenstates are orthogonal to each
other. And the Hamiltonian satisfies $\eta H=H^{\dagger }\eta $, which
implies the Hamiltonian is Hermitian under the new metric.

Under the new metric, the $S$-matrix elements are defined by%
\begin{equation}
S_{pk}=\,_{-}\left\langle \mathbf{N\theta }_{p}\right\vert \eta \left\vert
\mathbf{N\theta }_{k}\right\rangle _{+}=\,_{-}\left\langle \mathbf{N\theta }%
_{p}^{D}\left\vert \mathbf{N\theta }_{k}\right\rangle _{+}\right. .
\end{equation}%
The Lippmann-Schwinger formalism results in the $S$-matrix elements as%
\begin{equation}
S_{pk}=\delta _{kp}-2\pi i\delta (E_{k}-E_{p})T_{pk},
\end{equation}%
where
\begin{equation}
T_{pk}=\frac{g_{0}f_{p}\left\langle 0\right\vert V\left\vert \mathbf{N\theta
}_{k}\right\rangle _{+}}{\sqrt{2\omega _{p}\Omega }}=\frac{g_{0}f_{p}c_{k,+}%
}{\sqrt{2\omega _{p}\Omega }},
\end{equation}%
defines the transfer matrix, i.e., $T$-matrix, and the $S$-matrix element is
obtained as $S_{pk}=\delta _{kp}\exp (2i\delta _{k})$, where the phase shift
$\delta _{k}=\arctan \beta _{k}$ as $f_{k}\rightarrow 1$. Obviously, the $S$%
-matrix is unitary under the new metric. It is shown that the choice of the
above matric indeed does not the representation of the $S$-matrix and thus
the physically observable result remains the same as that in the
conventional matric.

\section{Conclusion}

Finally, let us briefly summarize our results as follows. For one boson mode
Lee model with an imaginary coupling constant, the Hamiltonian is
non-Hermitian with respect to the inner product defined in the conventional
QM, but we can use bi-orthogonal basis to find a proper inner product , so
that the Hamiltonian becomes Hermitian with respect to the new inner
product. This inner product insures the orthogonality and the positive
definiteness of norm of eigenstates. For the QFT Lee model with $g>g_{c}$,
the Hamiltonian is also non-Hermitian in the inner product defined in the
conventional QFT. Based on the bi-orthogonal basis approach the proper inner
product is also found. The proper inner product automatically insures the
Hermiticity of the Hamiltonian, the orthogonality and the positive definite
norms of the eigenstates. Physically the ghost state is killed in our
approach so that\ the unitarity of the $S$-matrix is recovered automatically.

Since the Lee model was found in 1954, there were many struggles
to find the proper metric so that the positive definite norms of
the eigenstates were insured. We have pointed out that the inner
product introduced by Pauli \textit{et al. }\cite{Pauli} did not
insure the positive norm of the ghost state. Our inner product is
so proper that the orthogonality and the positive definite norms
of the eigenstates are ensured simultaneously. And the Hamiltonian
becomes Hermitian in the Hilbert space with the new metric, which
ensures the unitary evolution of the quantum states. Our
investigations about the Lee\ model suggest that the Hermiticity
of the Hamiltonian is related to the definition of the inner
product space though this point is very clear in mathematics. For
the non-Hermitian Hamiltonian in one inner product space with the
metric, it may become Hermitian in another inner product space
with the different metric.

In the future work, we will used the bi-orthogonal basis to reconsider the
unitarity of the $S$-matrix in the higher sectors, such as $N\theta \theta $
sector \cite{Pauli, LW2, NTT1, NTT2, NTT3,NTT4}. We would establish the new
theoretical system of QFT by the bi-orthogonal basis, such as the new
Feynman rules, renormalization procedure \cite{LW1} and reduction formalism
\cite{LSZ,LSZ1,LSZ2}. The bi-orthogonal basis can also be used to reconsider
the strong-coupling\ induced unitarity problem for\ the generalized Lee
Model\ in quantum optics \cite{Zhou1,FanPRL,FanPRA,Zhou,Shi,Xu}, atom
physics \cite{F,Xu}, condensed matter physics \cite{A} and the high energy
physics \cite{neutron,LQCD}.

\appendix*

\section{Relationship between Lee mode and a \textquotedblleft Standard
model" in quantum optics}

The JC models with one mode and multi-modes can be regarded as a
\textquotedblleft standard model" in quantum optics or cavity QED, which
deals with the discrete levels interacting with some continuum \cite{L,S}.
In the appendix, we consider the relation between the Lee model and the
multi-mode JC model with the Hamiltonian%
\begin{eqnarray}
H_{\mathrm{sb}} &=&\Omega \left\vert e\right\rangle \left\langle
e\right\vert +\sum_{k}\omega _{k}a_{k}^{\dagger }a_{k}  \notag \\
&&+\sum_{k}G_{k}(a_{k}\left\vert e\right\rangle \left\langle g\right\vert
+h.c.),  \label{Hsb}
\end{eqnarray}%
where $\left\vert e\right\rangle $ and $\left\vert g\right\rangle $ are the
excited and ground states of a two level atom with energy level spacing $%
\Omega $. Here, $a_{k}$ ($a_{k}^{\dagger }$)\ denotes the annihilation
(creation) operator of the photon with dispersion relation $\omega _{k}=k=%
\sqrt{k_{x}^{2}+k_{y}^{2}+k_{z}^{2}}$. The coupling%
\begin{equation*}
G_{k}^{2}=\frac{\omega _{k}}{2\epsilon _{0}(2\pi )^{3}}D^{2}\cos ^{2}\varphi
_{k},
\end{equation*}%
of the atom to photon is determined by the angle $\varphi _{k}$ between the
atomic dipole momentum $D$ and the electric field polarization vector of
photon, where $\epsilon _{0}$ is dielectric constant in vacuum.

This model can be applied to describe the spontaneous emission of the atom.
In the single mode limit, the model becomes the single mode JC model%
\begin{eqnarray}
H_{\mathrm{JC}} &=&\Omega \left\vert e\right\rangle \left\langle
e\right\vert +\omega _{0}a^{\dagger }a  \notag \\
&&+G_{0}(a\left\vert e\right\rangle \left\langle g\right\vert +h.c.),
\end{eqnarray}%
which is just the one boson mode Lee model or QM Lee model discussed in Sec.
III.

Comparing Eq. (\ref{Hsb}) with the Lee model, we find that the multi-mode JC
model is formally the massless Lee model with $m_{N}=0$, $\mu =0$, $\Omega
=m_{0}$, and $G_{k}=g_{0}f_{k}/\sqrt{2\omega _{k}\Omega }$. However, the Lee
model has very different nature from that of the multi-mode JC model due to
the non-vanishing mass $\mu $. Thus, owing to the different dispersion
relation, the Lee model may has the stable $V$-particle state $\left\vert
\mathbf{V}\right\rangle $, but the multi-mode JC model possess the unstable
excited state that has the finite lifetime and decays to the ground state by
emitting photon.

In the following discussions, we calculate the lifetime of the excited state
of the two level atom by the scattering theory. Now, we do not limit the
form of the dispersion relation $\omega _{k}=w(k)$, but we assign it with a
lower cutoff $\omega _{\min }$ for any $k$. In this sense, we can generally
study how the form of dispersion relation effects on the existence of the
bound states.

We consider the scattering states%
\begin{equation}
\left\vert g,k\right\rangle _{\pm }=u_{k,\pm }\left\vert e\right\rangle
+\sum_{p}\chi _{k,\pm }(p)a_{p}^{\dagger }\left\vert g\right\rangle ,
\end{equation}%
corresponding to $\left\vert \mathbf{N}\theta _{k}\right\rangle _{\pm }$ in
the Lee model. Here,%
\begin{eqnarray}
u_{k,\pm } &=&G_{k}h_{sb,\pm }^{-1}(\omega _{k}),  \notag \\
\chi _{k,\pm }(p) &=&\delta _{kp}+\frac{G_{p}u_{k,\pm }}{\omega _{k}-\omega
_{p}\pm i0^{+}},
\end{eqnarray}%
are determined by the eigen-equation $H_{\mathrm{sb}}\left\vert
g,k\right\rangle _{\pm }=\omega _{k}\left\vert g,k\right\rangle _{\pm }$. We
define the function $h_{sb,\pm }(\omega )=\mathbf{G}_{\pm }^{-1}(\omega )$,
where%
\begin{equation}
\mathbf{G}_{\pm }(\omega )=[\omega -\Omega -\sum_{p}\frac{G_{p}^{2}}{\omega
-\omega _{p}\pm i0^{+}}]^{-1},  \label{hsb}
\end{equation}%
is just the retarded $\mathbf{G}_{+}(\omega )$ or advanced $\mathbf{G}%
_{-}(\omega )$ Green's function of the excited state. Then the $S$-matrix
elements $S_{kp}=$\ $_{-}\left\langle N\theta _{p}\left\vert N\theta
_{k}\right\rangle _{+}\right. $\ are%
\begin{equation}
S_{kp}=\delta _{kp}-2\pi i\delta (\omega _{k}-\omega _{p})G_{k}u_{k,+},
\label{SA}
\end{equation}%
and thus the scattering phase shift is%
\begin{equation}
e^{2i\delta _{k}}=\frac{\mathbf{G}_{sb,+}(\omega _{k})}{\mathbf{G}%
_{sb,+}^{\ast }(\omega _{k})}\text{.}  \label{ps}
\end{equation}

If there existed the bound states%
\begin{equation}
\left\vert B\right\rangle =\left\vert e\right\rangle +\sum_{p}\chi
_{B}(p)a_{p}^{\dagger }\left\vert g\right\rangle ,
\end{equation}%
in this system for $\chi _{B}(p)=G_{p}(E_{B}-\omega _{p})^{-1}$, the
energies $E_{B}<\omega _{\min }$ of the bound states $\left\vert
B\right\rangle $\ would satisfy%
\begin{equation}
h_{sb,\pm }(E_{B})=\text{Re}h_{sb,\pm }(E_{B})=0.  \label{eh}
\end{equation}%
Thus, we conclude that the poles of the $S$-matrix (\ref{SA})\ or phase
shift\ (\ref{ps}) determine the bound state energies.

Using Eq. (\ref{eh}), we discuss the existence of the bound states. Because $%
\partial _{\omega }h_{sb}(\omega )>0$ and $h_{sb}(\omega \rightarrow -\infty
)\rightarrow -\infty $, Eq. (\ref{eh}) possess the real roots $E_{B}$\ in
the regime $E_{B}<\omega _{\min }$ only if $h_{sb}(\omega _{\min }^{-})>0$,
where $\omega _{\min }^{-}=\omega _{\min }-0^{+}$. However, in the opposite
case $h_{sb}(\omega _{\min }^{-})<0$, i.e.,%
\begin{equation}
\omega _{\min }^{-}+\int d^{3}p\frac{G_{p}^{2}}{\omega _{p}-\omega _{\min
}^{-}}<\Omega ,  \label{O}
\end{equation}%
there is no real root in the regime $E_{B}<\omega _{\min }$. In this case (%
\ref{O}), Eq. (\ref{eh}) possess the imaginary solution $E_{\mathbf{quasi}%
}=E_{b}-i\gamma $\ in the regime $E_{b}>\omega _{\min }$. Though the
solution $E_{\mathbf{quasi}}$ is not the eigenvalue of the Hamiltonian due
to the Hermiticity of the Hamiltonian, it describes the unstable excited
state with the energy $E_{b}$ and the lifetime $1/\gamma $. For the small
decay rate, the real part $E_{b}$ and the decay rate $\gamma $ are
determined by%
\begin{equation}
\text{Re}h_{sb,+}(E_{b})=0,
\end{equation}%
and
\begin{equation}
\gamma =\frac{\text{Im}h_{sb,+}(E_{b})}{\left. \partial _{\omega }\text{Re}%
h_{sb,\pm }(\omega )\right\vert _{\omega =E_{b}}},
\end{equation}%
approximately \cite{ZS}. Here, Re$h_{sb,+}(E_{b})=$Re$h_{sb,-}(E_{b})$. On
the other hand, near the pole $E_{\mathbf{quasi}}$ the Green's function $%
\mathbf{G}_{+}(\omega )$ of the excited state is%
\begin{eqnarray}
\mathbf{G}_{+}(\omega ) &=&\frac{1}{\text{Re}h_{sb,\pm }(E_{b})+i\text{Im}%
h_{sb,\pm }(E_{b})},  \notag \\
&\sim &\frac{Z}{\omega -E_{b}+i\gamma },
\end{eqnarray}%
where $Z^{-1}=\left. \partial _{\omega }\text{Re}h_{sb,+}(\omega
)\right\vert _{\omega =E_{b}}$. It is remarkable that when the energy of
incident photon equals to $E_{b}$, the phase shift $\delta _{k}$ goes to $%
\pi /2$ so that the cross section tends to infinite, which exhibits a
typical resonance.

In the Lee model, $\omega _{\min }$ regardes the mass $\mu $ of boson, so
that under the condition $h(m_{N}+\mu )>0$ there always exists the bound
state $\left\vert \mathbf{V}\right\rangle $, i.e., the physical $V$-particle
state without the imaginary part of the eigen-energy. However, if $%
h(m_{N}+\mu )<0$ the bound state vanishes, and the stable $V$-particle state
becomes the unstable state. Thus, we conclude that in the Lee model, due to
the non-vanishing mass $\mu $, the system contains the bound state $%
\left\vert \mathbf{V}\right\rangle $ only if $h(m_{N}+\mu )>0$. The analytic
property of the Green's function $\mathbf{G}_{+}(\omega )$ in the whole
complex-$\omega $ plane is shown in Fig. 4, which exhibits that when $%
h(m_{N}+\mu )<0$, the bound state vanishes and two poles $E_{\pm }$\ in the
upper and lower planes emerge, the imaginary part $\gamma $\ of the $E_{+}$
is the decay rate of the unstable $V$-particle.
\begin{figure}[tbp]
\includegraphics[bb=16 502 588 730, width=9 cm, height=5 cm, clip]{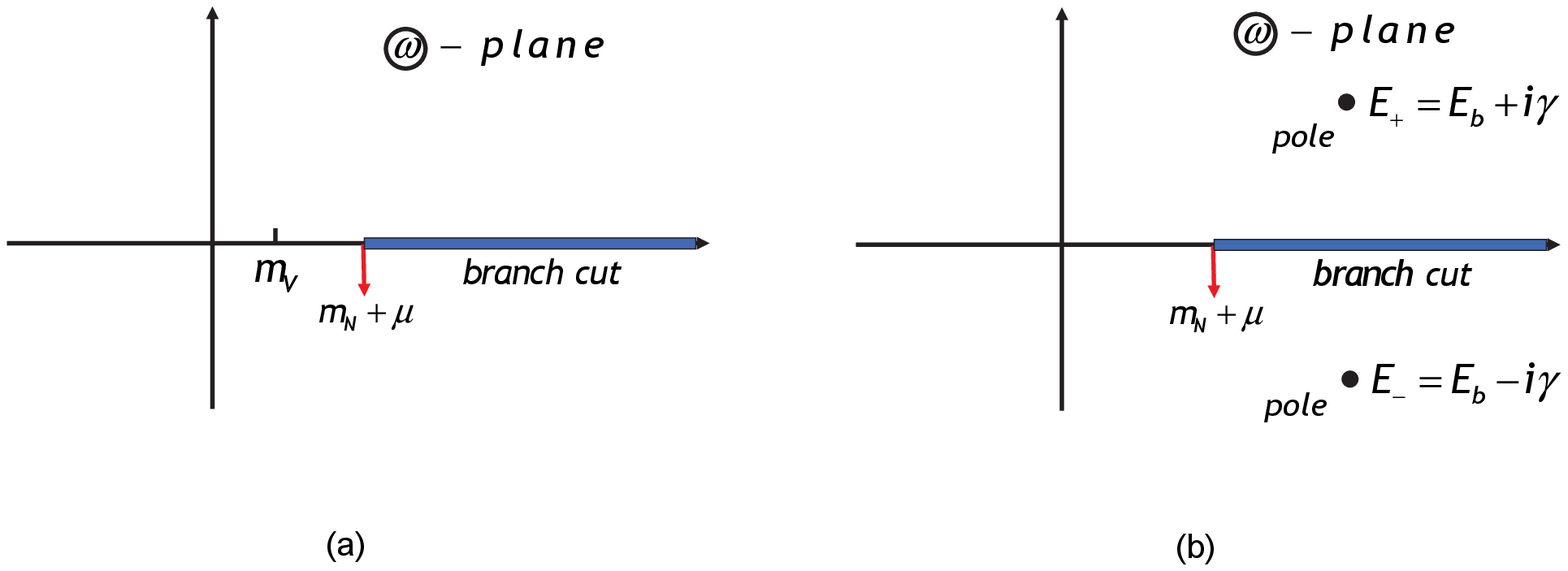}
\caption{(Color online) The analytic property of the Green's function $%
h^{-1}(\protect\omega )$ of JC model as an Lee model with vanishing mass of $%
\protect\theta $-particle: the Green's function possesses the branch cut in
the regime $\protect\omega >m_{N}+\protect\mu $, (a) if $h(m_{N}+\protect\mu %
)>0$, the Green's function has the pole $m_{V}$ corresponding to the bound
state energy; (b) if $h(m_{N}+\protect\mu )<0$, the bound state becomes the
unstable $V$-particle and two poles $E_{\pm }$\ in the upper and lower
planes emerge, where $\protect\gamma $ is the decay rate of the unstable $V$%
-particle \protect\cite{LW1}.}
\end{figure}
\begin{figure}[tbp]
\includegraphics[bb=37 279 568 606, width=8 cm, clip]{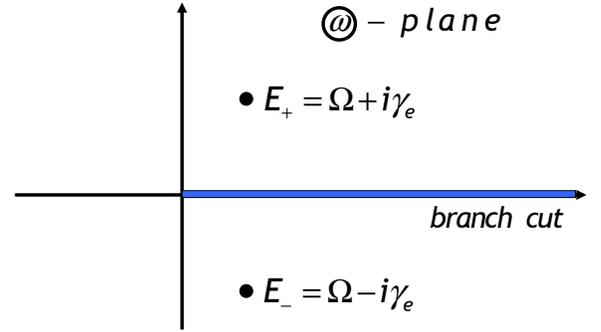}
\caption{(Color online) The analytic property of the Green's function $%
\mathbf{G}_{(}\protect\omega )$: because the vanishing mass $\protect\mu $,
the excited state energy $\Omega $ is embedded in the continuum $\protect%
\omega _{k}$ and the atom has the finite lifetime $1/\protect\gamma _{e}$.}
\end{figure}

For the multi-mode JC model, the mass of the photon is zero. We except the
excited state is greater than the ground state of the atom so that the
energy of the excited state always locates in the continuum of the photon
energy spectrum and the excited state is unstable with the lifetime $\tau
\sim 1/\gamma _{e}$. The renormalized energy of excited-state is$\ \Omega
_{r}=\Omega +\delta \Omega $, where the Lamb shift%
\begin{equation}
\delta \Omega =\mathcal{P}\int d^{3}p\frac{G_{p}^{2}}{\Omega _{r}-\omega _{p}%
}.
\end{equation}%
is determined by the condition Re$h_{sb,+}(\omega )=0$. When Lamb shift is
small, the decay rate of the excited state is%
\begin{equation}
\gamma \sim \pi \sum_{p}G_{p}^{2}\delta (\Omega -\omega _{p})=\frac{\Omega
^{3}D^{2}}{6\pi \epsilon _{0}},
\end{equation}%
which is just the spontaneous emission rate, which was given in many
references \cite{L,S,ZS}. The analytic property of the Green's function%
\begin{equation}
\mathbf{G}(\omega )=[\omega -\Omega -\sum_{p}\frac{G_{p}^{2}}{\omega -\omega
_{p}}]^{-1},
\end{equation}%
in the whole complex-$\omega $ plane is shown in Fig. 5, which exhibits that
the two poles $E_{\pm }$\ in the upper and lower planes emerge and the
imaginary part $\gamma $\ of the $E_{+}$ is the spontaneous emission rate.

\acknowledgments One (C. P. Sun) of the authors would like to thank Y. B.
Dai and C. Y. Zhu for many helpful discussion about the Lee model. The work
is supported by National Natural Science Foundation of China and the
National Fundamental Research Programs of China under Grant No. 10874091 and
No. 2006CB921205.


\begin{thebibliography}{99}
\bibitem{Lee} T. D. Lee, Phys. Rev. \textbf{95}, 1329 (1954).

\bibitem{Schweber} S. S. Schweber, \textit{An Introduction to Relativistic
Quantum Field Theory} (Row, Peterson and Co, Evanston, 1961), Chap. 12.

\bibitem{F} U. Fano, Phys. Rev. \textbf{124}, 1866 (1961).

\bibitem{A} P. W. Anderson, Phys. Rev. \textbf{124}, 41 (1961).

\bibitem{Zhou1} L. Zhou, F. M. Hu, J. Lu, and C. P. Sun, Phys. Rev. A,
\textbf{74}, 032102 (2006).

\bibitem{Redmond} P. J. Redmond, Phys. Rev. \textbf{112}, 1404 (1958).

\bibitem{EG-NL} G. Rasche and N. Straumann, Nuovo Cimento \textbf{4}, 4604
(1962).

\bibitem{Pauli} G. K\"{a}llen and W. Pauli, Dan. Mat. -Fys. Medd. \textbf{30}%
, 7 (1955).

\bibitem{LW1} T. D. Lee and G. C. Wick, Nucl. Phys. B \textbf{9}, 209 (1969).

\bibitem{LW2} T. D. Lee and G. C. Wick, Nucl. Phys. B \textbf{10}, 1 (1969).

\bibitem{Bender} C. M. Bender, S. F. Brandt, J. H. Chen, and Q. H. Wang,
Phys. Rev. D \textbf{71}, 025014 (2005).

\bibitem{Jones} H. F. Jones, Phys. Rev. D \textbf{77}, 065023 (2008).

\bibitem{Sun} C. P. Sun, Phys. Scr. \textbf{48}, 393 (1993).

\bibitem{Bi-O} P. T. Leung, W. M. Suen, C. P. Sun, and K. Young, Phys. Rev.
E \textbf{57}, 6101 (1998).

\bibitem{unstable} V. Glaser and G. K\"{a}llen, Nucl. Phys. \textbf{2}, 706
(1957).

\bibitem{Trubatch} S. L. Trubatch, Amer. J. Phys. \textbf{38}, 331 (1970).

\bibitem{JC} E. T. Jaynes and F. W. Cummings, \textit{Proc. IEEE }\textbf{51}%
, 89 (1963).

\bibitem{NTT1} R. D. Amado, Phys. Rev. \textbf{122}, 696 (1961).

\bibitem{NTT2} A. Pagnamento, J. Math. Phys. \textbf{6}, 955 (1965).

\bibitem{NTT3} A. Pagnamento, J. Math. Phys. \textbf{7}, 356 (1965).

\bibitem{NTT4} E. M. Kazes, J. Math. Phys. \textbf{6}, 1172 (1965).

\bibitem{LSZ} H. Lehmann, K. Symanzik, and W. Zimmermann, Nuovo Cimento
\textbf{1}, 1425 (1955).

\bibitem{LSZ1} M. S. Maxon and R. B. Curtis, Phys. Rev. \textbf{137}, B996
(1965).

\bibitem{LSZ2} M. S. Maxon, Phys. Rev. \textbf{149}, 1273 (1966).

\bibitem{FanPRL} J. T. Shen and S. Fan, Phys. Rev. Lett. \textbf{98}, 153003
(2007).

\bibitem{FanPRA} J. T. Shen and S. Fan, Phys. Rev. A \textbf{76}, 062709
(2007).

\bibitem{Zhou} L. Zhou, Z. R. Gong, Y. X. Liu, C. P. Sun and F. Nori, Phys.
Rev. Lett. \textbf{101}, 100501 (2008).

\bibitem{Shi} T. Shi and C. P. Sun, arXiv: quant-ph/0809.1279.

\bibitem{Xu} D. Z. Xu, H. Ian, T. Shi, H. Dong, and C. P. Sun, arXiv:
quant-ph/0812.0429.

\bibitem{neutron} C. C. Nishi and M. M. Guzzo, Phys. Rev. D \textbf{78},
033008 (2008).

\bibitem{LQCD} G. Z. Meng and C. Liu, Phys. Rev. D \textbf{78}, 074506
(2008).

\bibitem{L} W. H. Louisell, \textit{Quantum Statistical Properties of
Radiation} (Wiley, NewYork, 1973).

\bibitem{S} M. O. Scully and M. S. Zubairy, \textit{Quantum Optics}
(Cambridge University Press 1997).

\bibitem{ZS} S. R. Zhao, C. P. Sun, and W. X. Zhang, Phys. Lett. A \textbf{%
207}, 327 (1995).
\end{thebibliography}
\end{document}